\documentclass[pdflatex,sn-nature]{sn-jnl}

\usepackage{graphicx}%
\usepackage{multirow}%
\usepackage{amsmath,amssymb,amsfonts}%
\usepackage{amsthm}%
\usepackage{mathrsfs}%
\usepackage[title]{appendix}%
\usepackage{xcolor}%
\usepackage{textcomp}%
\usepackage{manyfoot}%
\usepackage{booktabs}%
\usepackage{algorithm}%
\usepackage{algorithmicx}%
\usepackage{algpseudocode}%
\usepackage{listings}%

\theoremstyle{thmstyleone}%
\theoremstyle{thmstyletwo}%

\theoremstyle{thmstylethree}%

\begin{document}

\title[Article Title]{Optoelectronic generative adversarial networks}


\author[1]{\fnm{Jumin} \sur{Qiu}}\email{qiujumin@email.ncu.edu.cn}
\equalcont{These authors contributed equally to this work.}

\author[1]{\fnm{Ganqing} \sur{Lu}}\email{402200220017@email.ncu.edu.cn}
\equalcont{These authors contributed equally to this work.}

\author*[2,3]{\fnm{Tingting} \sur{Liu}}\email{ttliu@ncu.edu.cn}

\author[1]{\fnm{Dejian} \sur{Zhang}}\email{dejianzhang@ncu.edu.cn}

\author[2,3]{\fnm{Shuyuan} \sur{Xiao}}\email{syxiao@ncu.edu.cn}

\author*[1]{\fnm{Tianbao} \sur{Yu}}\email{yutianbao@ncu.edu.cn}

\affil[1]{\orgdiv{School of Physics and Materials Science}, \orgname{Nanchang University}, \orgaddress{\city{Nanchang}, \postcode{330031}, \state{Jiangxi}, \country{China}}}

\affil[2]{\orgdiv{School of Information Engineering}, \orgname{Nanchang University}, \orgaddress{\city{Nanchang}, \postcode{330031}, \state{Jiangxi}, \country{China}}}

\affil[3]{\orgdiv{Institute for Advanced Study}, \orgname{Nanchang University}, \orgaddress{\city{Nanchang}, \postcode{330031}, \state{Jiangxi}, \country{China}}}


\abstract{Artificial intelligence generative content technology has experienced remarkable breakthroughs in recent years and is quietly leading a profound transformation. Diffractive optical networks provide a promising solution for implementing generative model with high-speed and low-power consumption. In this work, we present the implementation of a generative model on the optoelectronic computing architecture, based on generative adversarial network, which is called optoelectronic generative adversarial network. 
The network strategically distributes the generator and discriminator across the optical and electronic components, which are seamlessly integrated to leverage the unique strengths of each computing paradigm and take advantage of transfer learning. The network can efficiently and high-speed process the complex tasks involved in the training and inference of the generative model. 
The superior performance of these networks is verified by engaging three types of generative tasks, image generation, conditional generation, and image restoration. By synergistically combining the strengths of optical and electronic computing, the optoelectronic generative adversarial network paves the way for the development of more powerful and accessible artificial intelligence generative content technology that can unlock new creative possibilities across a wide range of applications.
}

\maketitle

\section{Introduction}\label{sec1}

Artificial intelligence generated content (AIGC) refers to content created or generated by artificial intelligence systems. Since ChatGPT was reported, AIGC has recently become not only a public word on social media, but also a hot spot in the academic field. Due to their outstanding generation ability, it has found many exciting applications in text\cite{10.5555/3295222.3295349}, images\cite{14,1}, and videos\cite{NEURIPS2022_39235c56}, revolutionizing the content creation process, Meanwhile, generative models have been applied to optimize the design of photonic devices\cite{Ma2021,SoRho+2019+1255+1261,https://doi.org/10.1002/adma.201901111}.
However, generative models often require tremendous computational resources to achieve optimal performance, and those foundation models even cost millions of dollars per day in electricity.

Optical computing, with its inherent parallelism and high-speed data processing capabilities\cite{Shen2017,Feldmann2019,PhysRevX.9.021032,Zhang2021,Liu2021,Wu:20,doi:10.1126/sciadv.abm2956}, may offer a potential solution to this problem. By leveraging the unique properties of light, optical computing systems could significantly improve the efficiency and scalability of AIGC models, enabling them to deliver high-quality results while consuming less energy and computational resource. In the pioneering work of Lin et al.\cite{doi:10.1126/science.aat8084}, diffractive optical network (DON, also known as diffractive deep neural network, D$^2$NN) was first proposed, which consisting of multilayer of three-dimensional printed diffractive optical elements operating at terahertz. The network performs inference and prediction at the speed of light through parallel computation and dense interconnection. Later, DONs have been effectively validated in performing specific inference functions\cite{Qian2020,Wang2023,doi:10.1126/sciadv.adg1505,Mengu2023,10.1117/1.AP.5.1.016003,https://doi.org/10.1002/adom.202200281,Goi2022,DuanChenLin+2023+893+903,LI2024110136,doi:10.1126/sciadv.adn2205}, such as image classification\cite{CHEN20211483,Liu2022,doi:10.1021/acsphotonics.0c01583,doi:10.1126/sciadv.abo6410}, saliency detection\cite{PhysRevLett.123.023901}, and decision-making\cite{Li:18}, and are implemented on various nanostructures\cite{Luo2022,https://doi.org/10.1002/lpor.202100732,10577112,doi:10.1126/science.adl1203,Chen2023,Cong}.
More recently, a reconfigurable DON based on optoelectronic fused computing architecture has been proposed\cite{Zhou2021,Xu2022}. By seamlessly combining optical and electronic components, this hybrid framework is poised to achieve extraordinary model complexity, with the ability to accommodate neural networks comprising millions of interconnected neurons.
Despite the remarkable progress of DONs in recent years, their primary applications have been largely limited to inference tasks, such as classification and recognition. To our knowledge, the development of generative models based on the DON architecture has not yet been achieved.

In this work, we implement generative model on the optoelectronic computing architecture, which is called optoelectronic generative adversarial network (OE-GAN). 
The networks build upon generative adversarial network (GAN)\cite{NIPS2014_5ca3e9b1}, combining the parallel processing capabilities of optical computing with the powerful representation learning of the GAN framework. 
By distributing the computational tasks of the generator and discriminator networks across optical and electronic components, the OE-GAN is able to achieve significant improvements in speed and energy efficiency compared to traditional all-electronic GAN implementations. 
This hybrid approach leverages the complementary strengths of optical and electronic  computation, while taking advantage of transfer learning, allowing the OE-GAN to generate high-quality results while overcoming the resource constraints that often limit the deployment of traditional generative models. 
The effectiveness of the proposed network is validated with three typical generative tasks, image generation, conditional generation, and image restoration.
As the field of optical computing continues to advance, we expect to see even more innovative applications of this technology in the context of AIGC.

\section{Results}\label{sec2}

\subsection{Network architecture}

\begin{figure*}[!ht]%
\centering
\includegraphics[width=\textwidth]{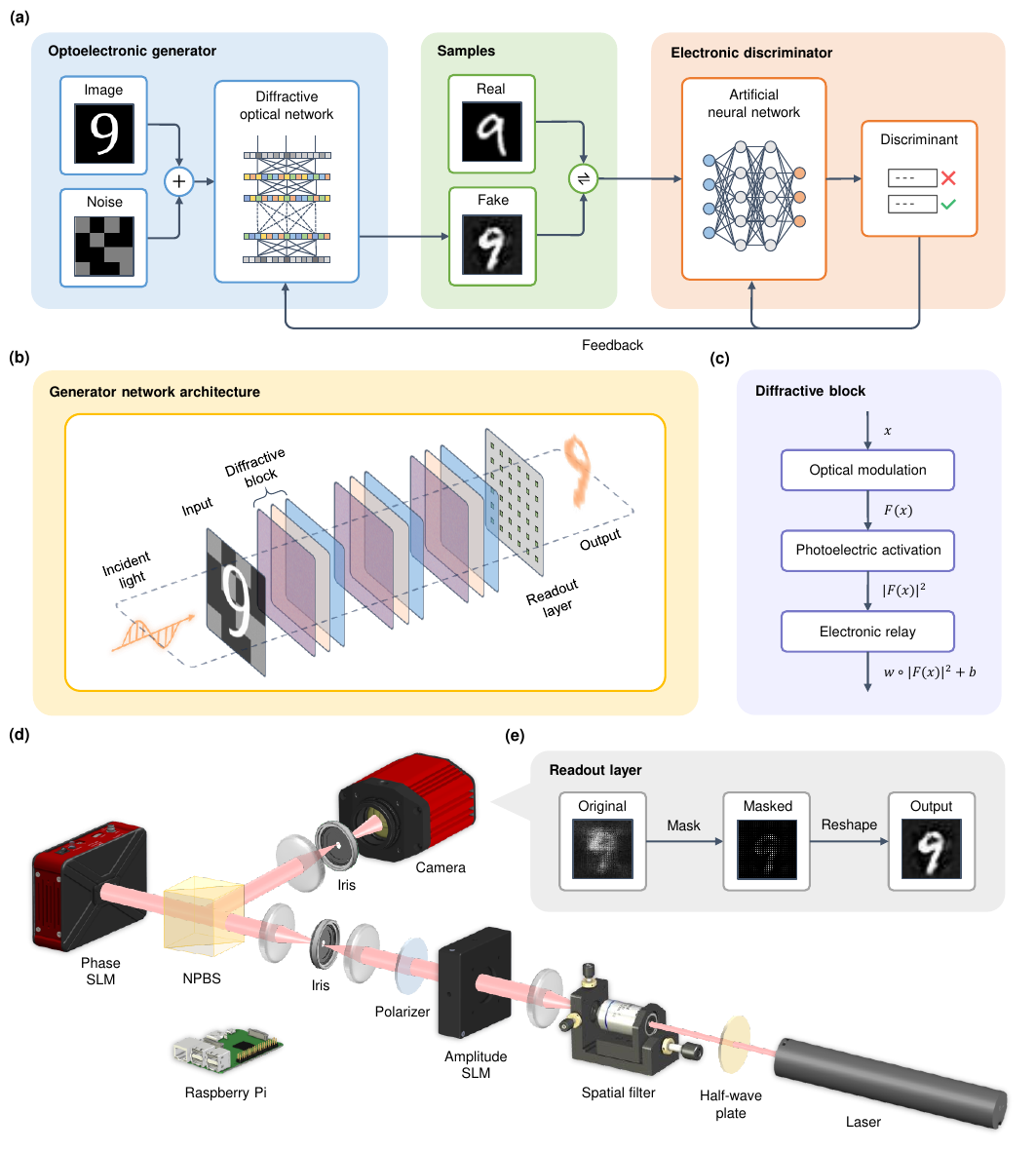}
\caption{\textbf{The network architecture of OE-GAN.} \textbf{a} The training framework of OE-GAN. The network learns to generate new data by pitting a optoelectronic generator against an electronic discriminator in an adversarial training process. \textbf{b} The architecture of the generator network. The input layer captures the image, a series of diffractive blocks allow the networks to learn and extract complex features and representations from the input data, and a readout layer maps the intensity distribution to generate the output result.  \textbf{c} The buliding block of network. \textbf{d} The experimental setup of OE-GAN. \textbf{e} The readout layer of network. The original output is masked and then reshaped to produce the result.}\label{fig1}
\end{figure*}

The network architecture of OE-GAN is illustrated in Fig. \ref{fig1}a. The system continues the fundamental philosophy of Wassertein GAN\cite{pmlr-v70-arjovsky17a}, helping to mitigate the vanishing gradient and mode collapse problems commonly observed in standard GANs, which is especially important in optical computing. The OE-GAN is composed of both optical and electronic components, seamlessly integrated to leverage the unique strengths of each computing paradigm. 
The optical computation of the OE-GAN system is responsible for performing high-speed, massively parallel operations, which are essential for the generative model's inference. 
Meanwhile, the electronic computation guides the optical computation model and thus trains the entire computation system.
It is important to note that once training is complete, only the optoelectronic generator performs the inference, which means that the vast majority of the computing process is optical, except for the dataflow control.

Specifically, OE-GAN consists of two neural networks: a DON-based optoelectronic generator and a electronic discriminator based on multilayer perceptron (MLP), which are trained in an adversarial manner. The core principle of the model is to pit these two networks against each other in a minimax game, where the goal is to find a Nash equilibrium. 
The generator network is responsible for generating synthetic data, that attempts to mimic the real data distribution. On the other hand, the discriminator network, is trained to distinguish between real and generated data samples.
During the training process, the generator network, based on the input noise and images, tries to generate data that can fool the discriminator into thinking it is real. The discriminator network then alternately receives real and fake samples, and tries to accurately identify the generated data as fake.
It is worth noting that the labels is optional during network training, thus OE-GAN allows unsupervised learning.

This adversarial training process continues until the generator network becomes skilled at producing realistic data that the discriminator can no longer reliably distinguish from the real data. The key to the success of network architecture is the competition between the generator and discriminator. As the generator becomes better at generating realistic data, the discriminator is forced to become more sophisticated in its ability to detect the generated data. This cycle of improvement continues, leading to the generator network producing increasingly convincing outputs.

The working principle of the optoelectronic generator is illustrated in Fig. \ref{fig1}b, which includes the specific free-space configuration. 
The network includes an input layer that takes in the noise and the image, encoded using an optical modulation device. The input is then processed through a series of stacked diffractive blocks that perform the optoelectronic computing. The optical signals are subsequently processed the results by a readout layer, allowing for the integration of optical and electronic subsystems. 

The diffractive block architecture is shown in Fig. \ref{fig1}c. First, the incident light with input information $x$ is modulated with a trained phase. This phase modulation is a key to the diffraction computation, which represents the parameters of the network. The diffraction computation, denoted as $F(x)$, is then performed as the light propagates through a certain distance. Then, the optical signals are converted to electronic form by the photoelectric effect that occurs at each pixel of the image sensor. This process effectively implements the activation function of the diffractive neurons, resulting in the output $\lvert F(x)\rvert^2$. 
These pixel are then respectively assigned weights and biases, as $w \circ\lvert F(x)\rvert^2 + b$, by electronic relay operations. In particular, this step is not strictly necessary for the core functionality of the diffractive block, and is primarily used to bridge the gap between the experimental implementation and the simulation models. In addition, since the operation is a Hadamard product, the computational cost is minimal.

After the series of diffractive computations, the output result can be obtained through a readout layer. The original output is first passed through a carefully designed spatial mask, which selectively retains only the relevant pixels. This masking operation helps to isolate the target information and discard any unwanted optical artifacts or background noise that may have been introduced during the diffractive computations. The masked output is then reshaped to produce the final result in the desired format. To further enhance the performance of the optoelectronic generator and mitigate experimental errors or discrepancies, in the actual experiment, we incorporate a small neural network as a post-processing step. This neural network can be trained to refine the output, improving the overall generation quality and consistency.

Once the training is complete, the target phase profiles of the diffractive layers are determined. We can realize the OE-GAN for generative based on existing optical devices.
Here, we choose an approach similar to the diffractive processing unit\cite{Zhou2021} to build the network, because of its reconfigurability and ability to support millions of neurons for computation. 
The experimental setup of the OE-GAN is shown in Fig. \ref{fig1}d. 
A laser beam is expanded using a microscope objective and lens, while a half-wave plate can be embedded to adjust the polarization direction.
The amplitude spatial light modulator (SLM) for optical encoding the input of the network or each diffractive block, followed by a linear polarizer and two relay lenses to adjust the polarization direction and size. 
The light is projected onto the phase SLM for phase modulation, and the diffraction pattern is imaged onto the camera. 
After that, the output image is input to the amplitude SLM for the next diffractive block until the end of the network computation. 
These light modulation devices are very fast, allowing real-time computation.  
We use a Raspberry Pi to control the above devices, and perform electronic relay and readout layer operations.

\subsection{Image generation}

In our first implementation, we perform a OE-GAN to generate images, we use the MNIST and Fashion-MNIST dataset as the real sample distribution. The input of network is a $4 \times 4$ random binary matrix image, which serves as the input noise. This low-dimensional input is then transformed through a series of operations within the optoelectronic generator, to produce plausible, high-quality digit samples. Since the samples do not contain any labels, the training of the OE-GAN is performed with unsupervised learning. The network learns to capture the underlying patterns and structures of the real digit samples without the need for explicit labels or supervision.

\begin{figure*}[htbp]%
\centering
\includegraphics[width=\textwidth]{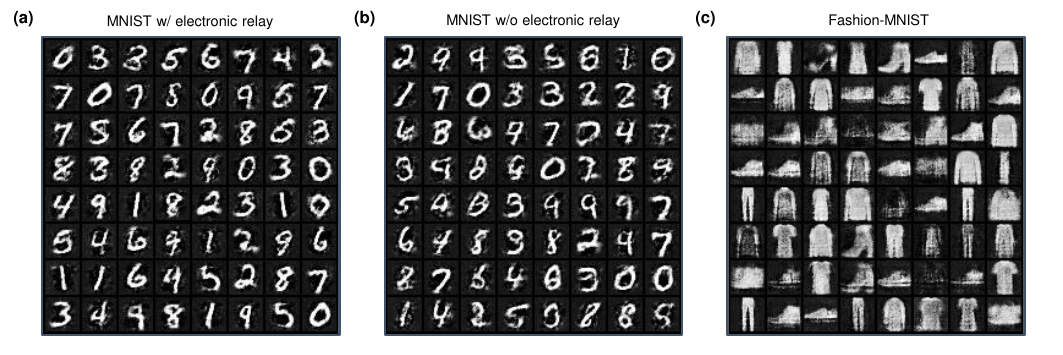}
\caption{\textbf{OE-GAN for image generation.} \textbf{a,b} Comparison of MNIST image generation results with and without electron relay. \textbf{c} The image generation results of Fashion-MNIST.}\label{fig2}
\end{figure*}

Fig. \ref{fig2}a,b show a batch of results of the MNIST generation, it can be seen that the majority of the generated samples demonstrate a remarkable level of detail and fidelity, closely mimicking the characteristics of the real handwritten digits from the MNIST dataset. 
It is worth noting that most of the samples generated by the network without electronic relays are equally effective compared to those generated with the additional processing step. However, a small number of the generated samples do not meet the same high standard of quality, suggesting that there is still room for improvement in the network's ability to consistently produce perfect results.
This results indicate that the electronic relay can improve the performance of the network to some extent. 
Besides that, the electronic relay operations are primarily introduced to bridge the gap between the experimental implementation and the simulation models. 
In principle, the electronic relay component may not be an absolute necessity under simulation conditions, and could potentially be optimized or even eliminated in future iterations of the experimental design of OE-GAN to reduce the ratio of electrical computation.

Fig. \ref{fig2}c shown the clothing samples generated based on Fashion-MNIST. This dataset is more challenging than the binarized handwritten digits of the MNIST dataset, because the Fashion-MNIST samples are more complex grayscale images of different clothing items. Despite the increased complexity, the OE-GAN is able to successfully capture and reproduce the salient features of the clothing samples. The generated images clearly depict the contours and gray tones of the different categories of clothing, demonstrating the remarkable generative capabilities of the optoelectronic architecture.

\begin{figure}[htbp]%
\centering
\includegraphics[width=0.5\textwidth]{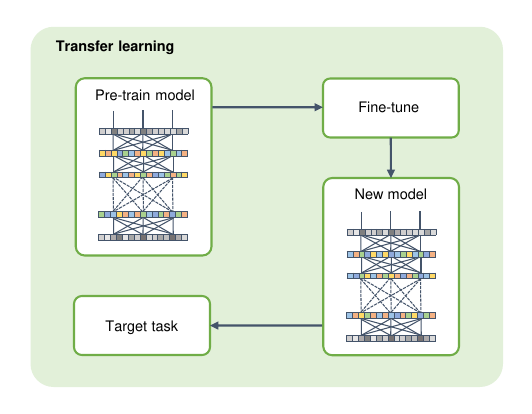}
\caption{\textbf{Transfer learning for OE-GAN.} Transfer learning leverages the knowledge gained from solving one problem and applies it to a different but related problem.
By taking advantage of pre-train model and fine-tuning it on the target task, transfer learning can reduce training time and improve performance.}\label{fig2-1}
\end{figure}

In addition, it is important to note that by leveraging the knowledge and representations learned during the initial training on image generation tasks, we can use the pre-trained model as a powerful starting point for subsequent implementations, as shown in Fig. \ref{fig2-1}. Specifically, we use the learned phase profile from the image generation task as the pre-training model. In the training of subsequent tasks, the initial phase profile of the generator network is the pre-training model. Due to the high relevance of these tasks in this work, the approach can significantly reduce the time and resources required to adapt the OE-GAN to new data domains and tasks, since the network already has a strong foundation, see Supplementary Fig. S1 and S2 for details. Such transfer learning capabilities are particularly valuable in the context of optical computing, where the design and optimization of the physical components can be a complex and time-consuming process. By reusing the pre-trained model, we can focus their efforts on fine-tuning and integrating the network with the specific optoelectronic hardware, rather than having to rebuild the entire system from scratch.

\subsection{Conditional generation}

\begin{figure*}[htbp]%
\centering
\includegraphics[width=\textwidth]{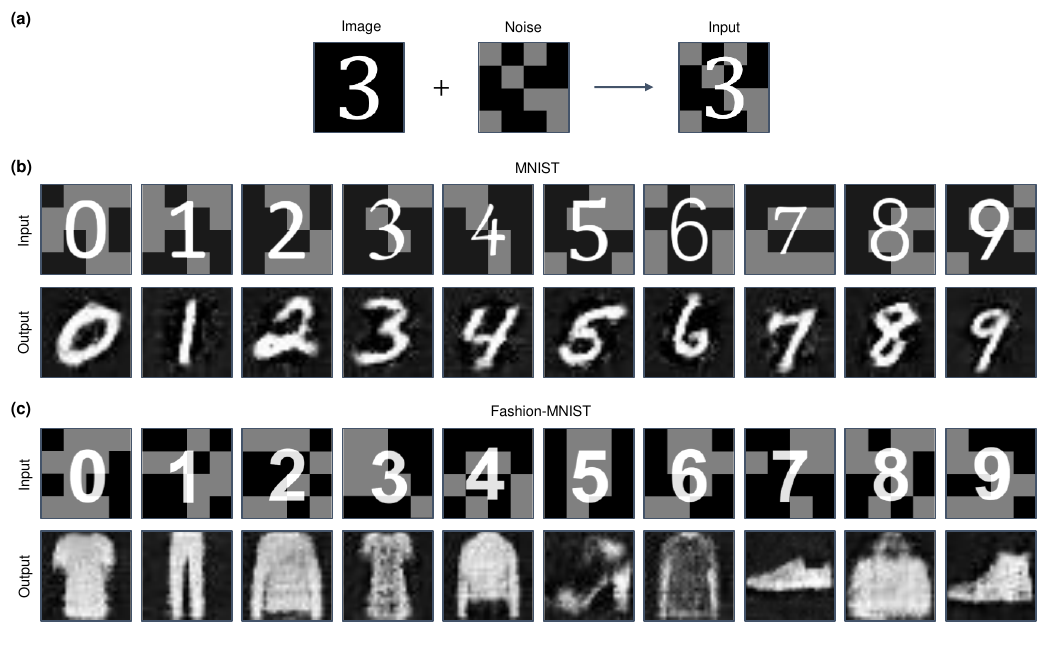}
\caption{\textbf{OE-GAN for conditional generation.} \textbf{a} The processing of input image. \textbf{b,c} The conditional generation results of MNIST and Fashion-MNIST.}\label{fig3}
\end{figure*}

In our second implementation, we build a conditional OE-GAN that can generate images conditioned on a given class. The introduction of conditional generation capabilities allows the OE-GAN to produce tailored samples that adhere to predefined constraints, such as the target digit class. This added control can be highly valuable in various applications, where the ability to generate class-specific samples is crucial. As shown in Fig. \ref{fig3}a, we use the sum of the conditional image and the noise as the input image, the network is now able to generate conditioned on a specified class. The conditional image is a number that represents the desired class of generated image. In addition, for handwritten digits generation, these numbers have different fonts. The different fonts add some challenges to the network, as the network must now learn to generalize across a diverse set of font styles and variations. 

The results are shown in Fig. \ref{fig3}b,c, generated by transfer learning based on previous implementation, and Supplementary Fig. S3 shows a batch of the conditional generation results. This approach is effective in accelerating the speed of training and improving the quality of generation.
As evidenced by the generated samples, the OE-GAN has successfully learned to generate images that accurately reflect the specified target classes. The visual quality of the output samples is as good as unconditional generation, indicating that the introduction of this conditional mechanism does not degrade the performance of the network. In additon, despite the potential challenges posed by the diverse font representations used as conditional inputs, the OE-GAN has managed to capture the nuanced characteristics of each class and translate them into satisfactory generated samples.

\subsection{Image restoration}

\begin{figure*}[htbp]%
\centering
\includegraphics[width=\textwidth]{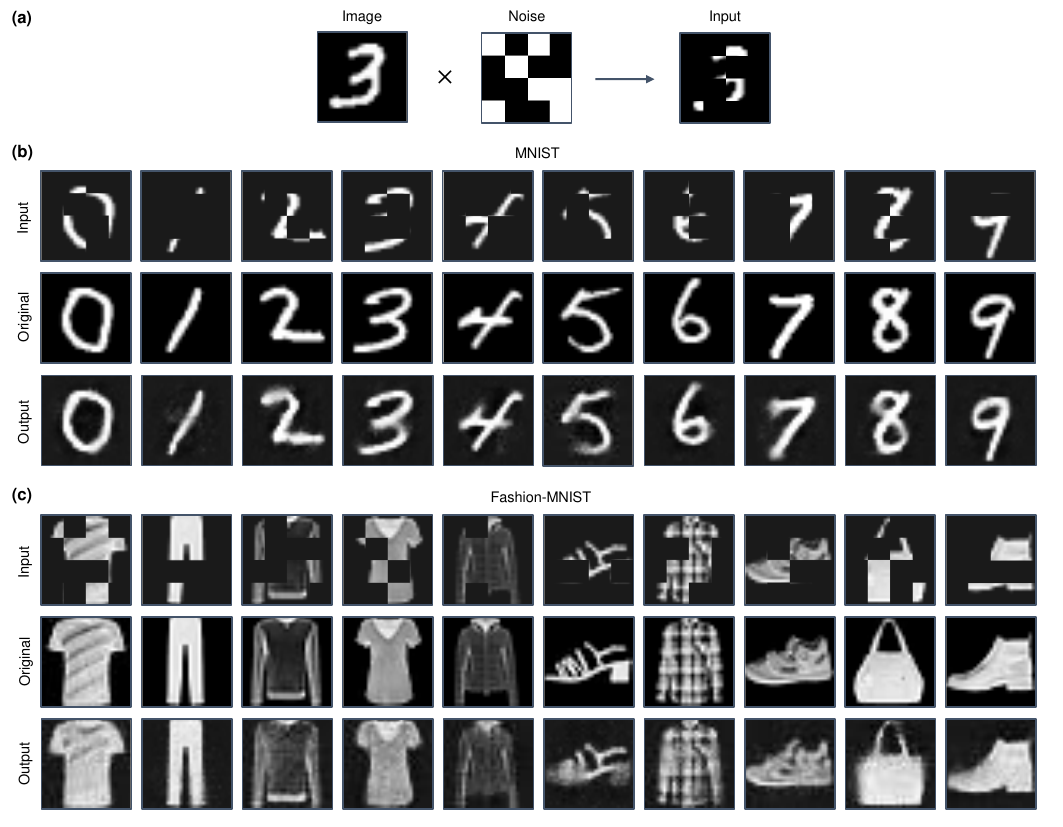}
\caption{\textbf{OE-GAN for image restoration.} \textbf{a} The processing of input image. \textbf{b,c} The image restoration results of MNIST and Fashion-MNIST.}\label{fig4}
\end{figure*}

In our third implementation, we demonstrate the capability of OE-GAN for image restoration, which is a typical application of generative model. The challenges of the network is to recover high-quality images from heavily degraded inputs. As shown in Fig. \ref{fig4}a, we multiply the noise and the input image to get the masked image, simulating the situation where data integrity compromised. The network must now learn to extract meaningful information from the partially obscured input and reconstruct the original image with a high degree of fidelity, demonstrating its strength in tackling challenging image restoration tasks.

The successful restoration of the masked images in both MNIST and Fashion-MNIST, and is highly similar to the original image, as the results presented in Fig. \ref{fig4}b,c. In addition, a batch of image restoration results is shown in Supplementary Fig. S4. This will undoubtedly showcase the sophisticated problem-solving capabilities of the OE-GAN model. The training of this network is also based on a pre-trained model. The network's capacity to recover high-quality, visually-coherent images from heavily degraded inputs highlights its potential practical applications of the OE-GAN architecture beyond mere image generation. Compared to previous research\cite{Rahman2023}, this approach does not require introducing additional neural network encoders, and is able to recover images under random positional occlusion.

\subsection{Experimental demonstration}

\begin{figure*}[htbp]%
\centering
\includegraphics[width=\textwidth]{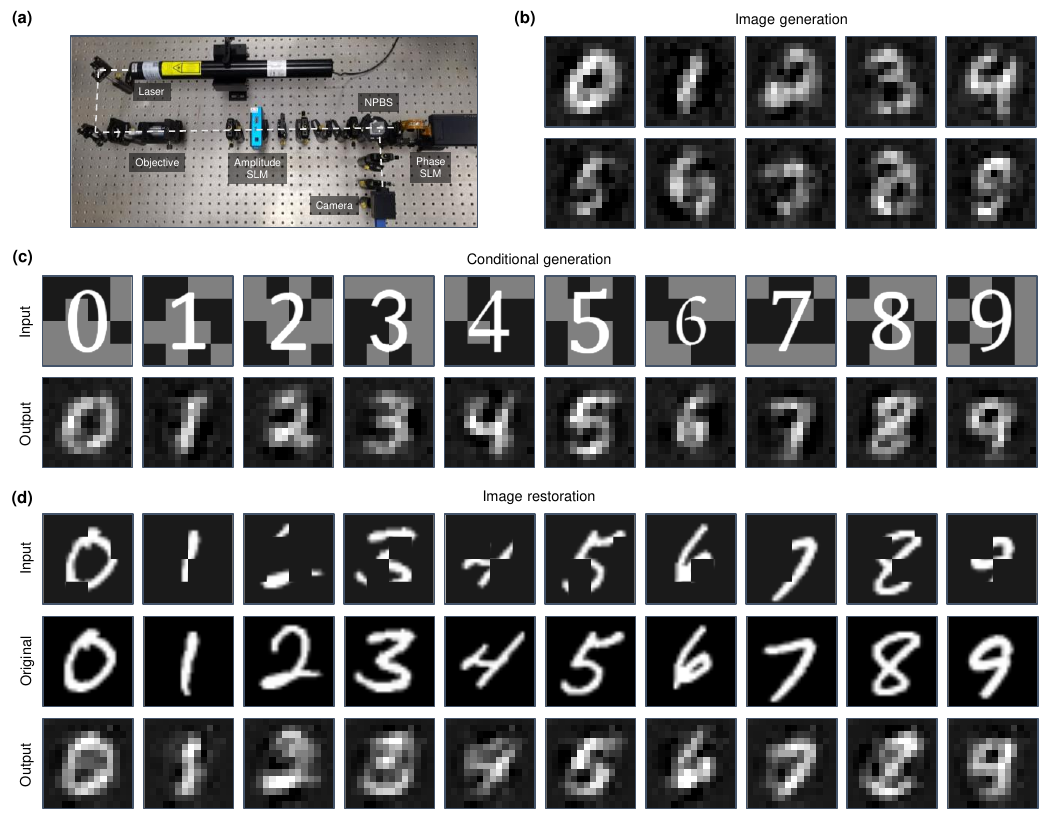}
\caption{\textbf{Experimental demonstration of OE-GAN.} \textbf{a} The photo of the experimental system. \textbf{b--d} The experimental results of image generation, conditional generation, and image restoration, respectively.}\label{fig5}
\end{figure*}

Finally, to evaluate the real experimental performance of the OE-GAN, we built an experimental system using off-the-shelf optical modulation devices, as shown in Fig. \ref{fig5}a. 
The experimental validation on the three key implementations previously demonstrated in simulation: the MNIST-based image generation, conditional generation, and image restoration. It ensures that the real-world performance of the OE-GAN is thoroughly tested on a variety of tasks, providing a comprehensive evaluation of the network's strengths and potential limitations.

Recognizing the inherent challenges of working with physical hardware, we reduce the output resolution of the experimental results to $14 \times 14$, as compared to the higher-fidelity simulations. This pragmatic approach helps to mitigate the potential impact of experimental errors or discrepancies, and ensures that the evaluation remains grounded in realistic conditions and constraints.

The experimental results presented in Fig. \ref{fig5}b--d provide a compelling evidence of the actual capability of the OE-GAN model, the network consistently achieves acceptable results, still generating the contours and details of the digits, although not as good as  the simulations. The network successfully generated images according to the requirements. Despite the transition from simulation to real-world implementation, the network has demonstrated its ability to maintain generation performance in these tasks. The successful transfer of the OE-GAN's capabilities to the experimental setup underscores the network's potential for practical deployment in real-world applications. 

\section{Discussion}\label{sec13}

We have demonstrated the OE-GAN for generative tasks. The network seamlessly integrates optical and electronic components to harness the strengths of each computing paradigm for high-speed and energy-efficient generative content creation. The optical computation handles the massively parallel and high-speed operations essential for generating synthetic data, while the electronic computation guides the optical model and manages the overall training process. This hybrid approach enables the OE-GAN to overcome the resource constraints that often limit the deployment of traditional all-electronic generative models. 
In addition, by leveraging transfer learning and pre-trained model, the OE-GAN can benefit from the knowledge and representations acquired in related tasks, significantly reducing the time and resources required for training from scratch.

We believe that this work represents an important step towards the development of more powerful and accessible AIGC technology.
While preliminary, this research suggests that the DON has great potential for processing complex generative tasks. The fusion of optoelectronic computing and generative model could provide a promising way to bridge the bridging the gap between the growing demand for AIGC and the resource constraints.

\section{Methods}\label{sec11}

\subsection{Experimental setup}
A He-Ne laser (HNL100L, Thorlabs) with a working wavelength of 632.8 nm is used as the light source. An amplitude SLM (TSLM023-A, CAS Microstar) with a pixel size of 26 $\mu$m and a resolution of 1,024$\times$768, which can modulate the incident light to display 8-bit grayscale images. A phase SLM (PLUTO-2.1, HOLOEYE Photonics) with a pixel size of 8 $\mu$m and a resolution of 1,920$\times$1,080 is used to modulate the phase of the wavefront. A grayscale CMOS camera (GS3-U3-51S5M, Teledyne FLIR) with a resolution of 2,448$\times$2,048 and a pixel size of 3.45 $\mu$m is used to capture the output. A Raspberry Pi 5 control the above devices by self-developed Python scripts.

\subsection{Network details}

We use the open-source machine learning framework Pytorch to build the algorithm. The training process is implemented using Python (v3.10.10) and Pytorch (v2.0.0) framework on a desktop computer with Intel Core i7-13700K CPU, Nvidia GeForce RTX 4070 Ti GPU, and 16 GB RAM. 

We train the network individually for each task, the structure of each network is identical. The networks use RMSprop optimization parameters, the learning rate is $10^{-3}$ for the generator and $10^{-4}$ for the discriminator. In addition, we clip the input image to be between $(0.1, 1)$, which makes the training more stable. 

The generator network consists of three diffractive blocks, and each layer has $1,080 \times 1,080$ neurons. Due to the differences in resolution between various devices, the input to each block is downsampled to $310 \times 310$, which can be analogized to the pooling operation. The electronic relay contains the same weights and biases as the number of neurons. In the simulation, the readout layer contains $28 \times 28$ equally spaced masks, each mask size is $5 \times 5$. We tested the effect of different mask sizes on the results, as shown in the Supplementary Fig. S5. In the experiments, the number of masks in the readout layer is $14 \times 14$, and a fully connected layer of the same size is included. In addition, we implement the adaptive training to optimize the phase profile to calibrate the misalignment error of the system.

The discriminator network consists of three fully connected layers, the first layer maps the input $28 \times 28$ image to 512 features, the second layer maps 512 features to 256 features, and the third layer maps 256 features to one feature. Rectified linear unit (ReLU) activation is applied after the first two layers and the final output is generated by the third layer.

\subsection{Loss functions}

The loss function of OE-GAN is essentially the same as that of Wasserstein GAN, which is derived from the Wasserstein distance between the generated distribution and the real data distribution, but the loss function is slightly different for each task. The key components of the loss function are generator loss and discriminator loss. Specifically, the generator loss aims to minimize the Wasserstein distance between the generated distribution and the real data distribution. It is formulated as:
\begin{equation}
L_G=-\mathbb{E}[D(G(z))],
\end{equation}
where $G(z)$ represents the generated samples, and $D(.)$ is the discriminator network.

The discriminator loss seeks to maximize the Wasserstein distance between the real data distribution and the generated distribution. It is formulated as:
\begin{equation}
L_D=-\mathbb{E}[D(x)]+\mathbb{E}[D(G(z))],
\end{equation}
where $x$ represents the real samples.

During the training process, the generator parameters are first fixed and the discriminator parameters are updated according to the $L_D$, and the parameters are clipped to be between $(-0.01, 0.01)$. Then, the discriminator is fixed and the generator parameters are updated according to the $L_G$, to drive the generated samples closer to the real data samples. The above process is then continuously looped until convergence. In addition, for image restoration task, the loss function of generator is the weight sum of $L_G$ and mean square error (MSE) of $x$ and $G(z)$ to guide the generated image is as close as possible to the original.

\section*{Data availability}

All data are available in the main text and the Supplementary Information.

\section*{Code availability}

The codes used in this work is available in the GitHub repository at \url{https://github.com/qiujumin/OE-GAN}.



\section*{Acknowledgments}

This work was supported by the National Natural Science Foundation of China (Grant Nos. 12064025, 12264028, 12364045, and 12304420), the Natural Science Foundation of Jiangxi Province (Grant Nos. 20212ACB202006, 20232BAB201040, and 20232BAB211025), and Young Elite Scientists Sponsorship Program by JXAST (2023QT11, 2025QT04).

\section*{Author Contributions}

J.Q. and G.L. conceived the idea. J.Q., T.L, and S.X. performed the theoretical calculation and numerical simulations. G.L. and D.Z. performed the experiments. J.Q., T.L., and T.Y. prepared the manuscript. T.L. and T.Y. supervised the project. All authors discussed the results and prepared the manuscript.

\section*{Competing Interests}

The authors declare no competing interests.

\end{document}